\documentstyle[11pt,twoside,paspconf,epsf]{article}

\begin{document}

\title{The X-ray Behavior of Two CVs: TT Ari and DP Leo}

\author{Craig R. Robinson and France A. C\'ordova}

\affil{Department of Astronomy and Astrophysics,
The Pennsylvania State University, 525 Davey Laboratory, University Park,
PA~16802, USA}

\begin{abstract}
We present ROSAT PSPC observations of the nova-like, or intermediate polar, TT
Ari and the eclipsing polar DP Leo.
Observations of TT Ari were performed as part of a simultaneous multiwavelength
campaign.  The X-ray
spectrum from the ROSAT PSPC \mbox{(0.2-2 keV)} was combined with Ginga
observations (2-37 keV) to suggest the presence of at least
three distinct emission components: an optically thin plasma
($kT_{RS}$=$0.76^{+0.20}_{-0.16}$ keV), a dominating bremsstrahlung
($kT_{br}$=$16.8^{+4.3}_{-3.4}$ keV)
continuum and one or more emission lines fit with a single gaussian centered at
6.57$\pm$0.18 keV.  IUE observations show a modulation
in the equivalent width of the CIV ($\lambda1549$) absorption profile on the
spectroscopic period indicating a complex wind structure.

The highly variable X-ray light curve
of DP Leo exhibits an intensity dip prior to eclipse which has not previously
been
observed.  The dip is interpreted as the eclipse of the main accretion
region by an accretion stream varying in shape or impact position on the white dwarf
with time.  The soft X-ray spectrum is well fit
by either a blackbody (kT = $24.8^{+2.6}_{-8.1}$ eV) or a soft power law
spectrum ($\alpha \sim$ 4.0).  No evidence exists for accretion
onto the stronger magnetic pole and severe limits are placed upon the flux from
any hard bremsstrahlung component. A distance upper limit of 500 pc is obtained
from the X-ray absorption and a distance estimate of 450 pc was
derived through the use of published photometry.
\end {abstract}

\section{Introduction}
Disk formation occurs in many cataclysmic variables (CVs) and numerous other
astrophysical objects from proto-planetary systems to active galactic nuclei.
The close proximity, high space density and peak disk luminosity at easily
observable wavelengths allow for numerous studies of CV disks and the
application of the knowledge
gained to models of disks elsewhere in the Universe.  However, CVs also provide
laboratories for studying
the effect on disks of varying mass accretion, tidal forces and magnetic
fields.
The disruption and distortions of disks and accretion streams in the presence
of a magnetic field may be studied through observations of the
magnetic CV systems known as intermediate polars (IPs) and polars.

We present ROSAT PSPC observations of a possible IP, TT Ari, and the eclipsing
polar DP Leo.  The analysis
presented here is part of a larger study on the sources and structure of
magnetism in CVs.

\section{TT Arietis}
TT Arietis is an unusual CV exhibiting several distinct periodicities
in optical photometry, similar to IPs, yet over long time scales it
occasionally drops into low states characterized by the ``anti-dwarf nova'' VY
Scl
systems (Shafter et al. 1985).
The system's orbital period is observed from radial velocity measurements of
emission
lines to be 0.13755114(13) d (Thorstensen et al.\ 1985) while the photometric
period is shorter with a value around 0.1329 d (see Tremko 1992). This binary
was found to be a hard X-ray source using the Einstein X-ray
satellite (C\'ordova, Mason \& Nelson 1981).  The Einstein data
are consistent with modulations of the X-ray flux with either the spectroscopic
or
photometric periods (Jensen et al.\ 1983) while EXOSAT data exhibit no
conclusive
evidence for strictly periodic modulations (Hudec et al.\ 1987). A 4 day
periodicity
found in U-band photometry (Semeniuk et al.\ 1987) suggests the presence of
variations at the beat period between the spectroscopic period ($\sim$ 3.3 hr)
and the photometric
period ($\sim$ 3.2 hr). This behavior is similar to the intermediate polar TV
Col
where the spectroscopic period ($\sim$ 5.5 hr) is slightly larger than the
photometric
period ($\sim$ 5.2 hr) and a variation at the 4 day beat period is observed
(Hellier et al.\ 1991).

The TT Ari observations presented here are part of a simultaneous
multiwavelength campaign
combining observations from ROSAT, IUE, Ginga and several ground-based
observatories. Future papers will explore the observations across all of these
wavelengths.
The ROSAT PSPC observations of TT Ari occurred 1991 August 1.3 to 2.2
for a total of approximately 25 ksec of usable data.  Ginga and IUE
observations were obtained
over part of the ROSAT observations.
The background subtracted PSPC count rate was
0.42$\pm$0.01 cts $\rm{s^{-1}}$.

\begin{figure}[htbp]
\plotfiddle{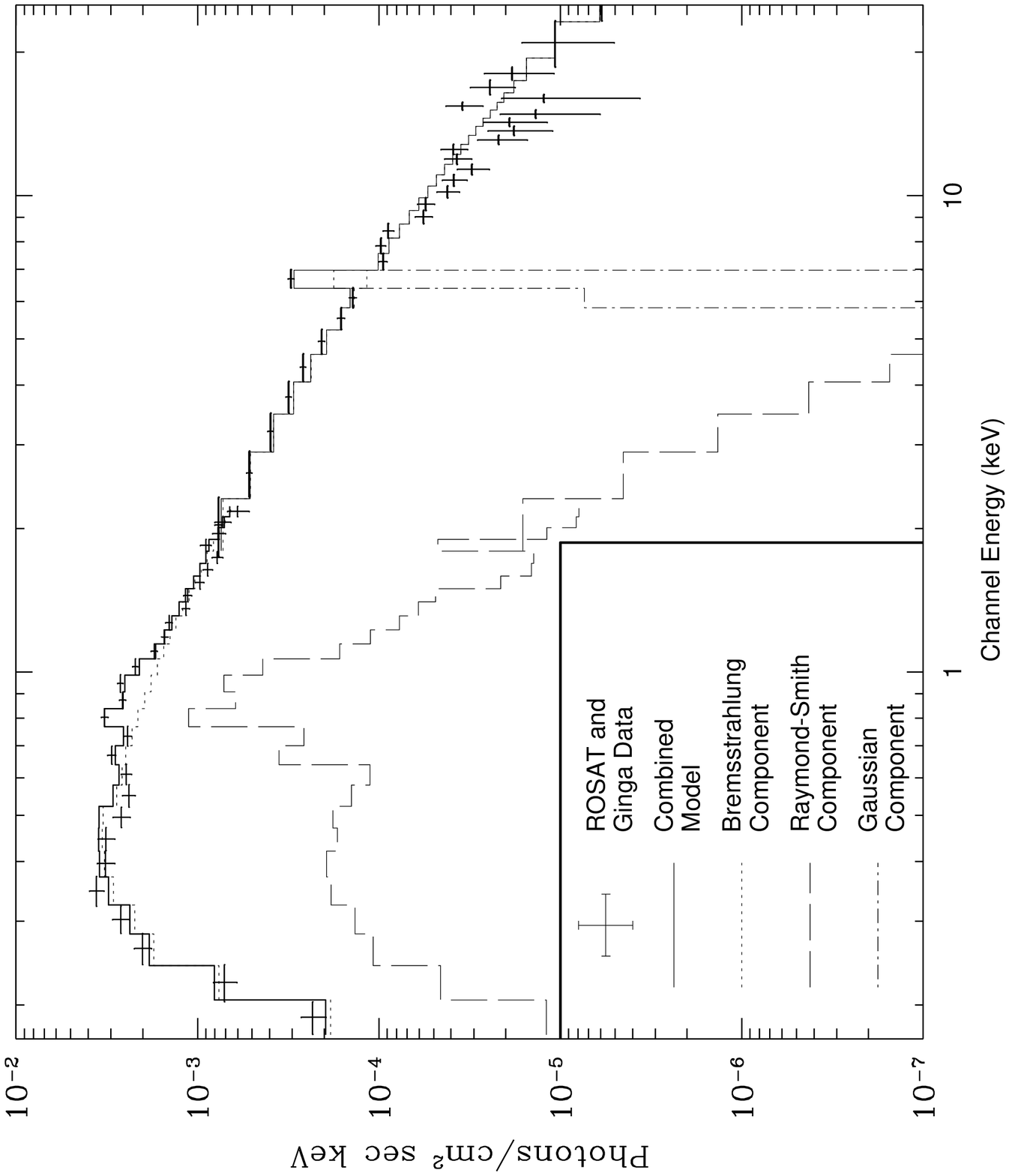}{8.8cm}{270}{50}{50}{-210}{280}
\caption{The Ginga and ROSAT X-ray spectra of TT Ari.}
\label{fig-1}
\plotfiddle{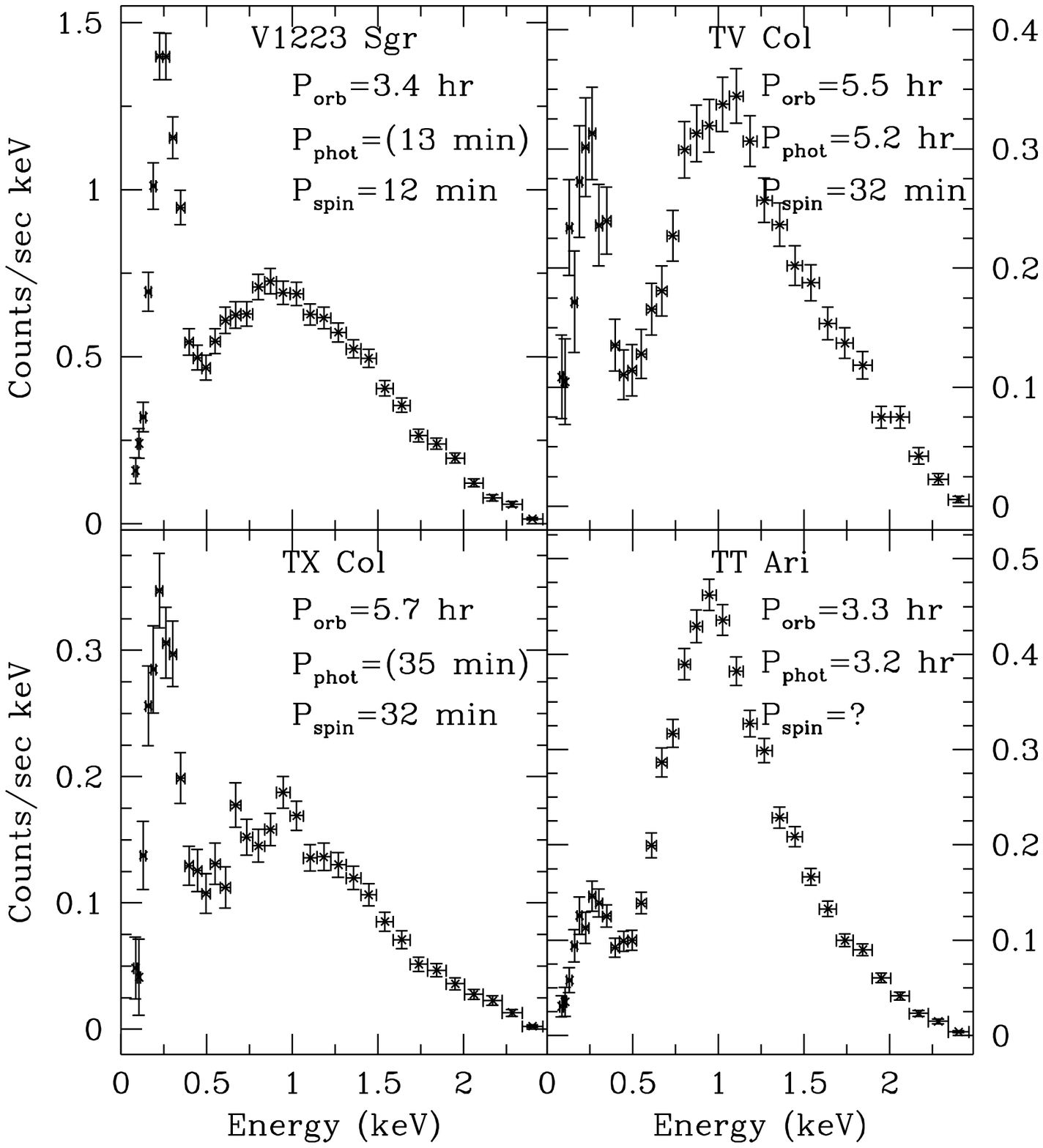}{9.4cm}{0}{60}{55}{-190}{-100}
\caption{ROSAT count spectra of 3 IPs and TT Ari.}
\label{fig-2}
\end{figure}

Spectral models were fit simultaneously to the ROSAT and Ginga data sets.
These observations are described
more fully within Robinson, C\'ordova \& Ishida (1993).  Figure 1 shows the
photon spectrum resulting
from the best fit model together with the observations.  The spectrum is
dominated by a thermal bremsstrahlung
continuum characterized by a temperature of kT = $16.8^{+4.3}_{-3.4}$ keV.  An
additional component was modeled as a Raymond-Smith plasma of temperature
$0.76^{+.20}_{-.16}$ keV.  The derived absorption column density was determined
to be $\rm{N_H}$ = 4.4$\pm$0.5 $\times 10^{20}$ $\rm{cm^{-2}}$.  An emission
line centered at 6.57$\pm$0.18 keV,
with an assumed line width ($\sigma$) of 0.1 keV and resulting equivalent width
of 0.89 keV, was added to the other components.  A strong iron (K$\alpha$?)
emission line was suggested by Ishida (1991) in the Ginga spectra of all
observed polars and IPs.  Other iron lines may exist but cannot be
distinguished by Ginga.

\begin{figure}[hb]
\plotfiddle{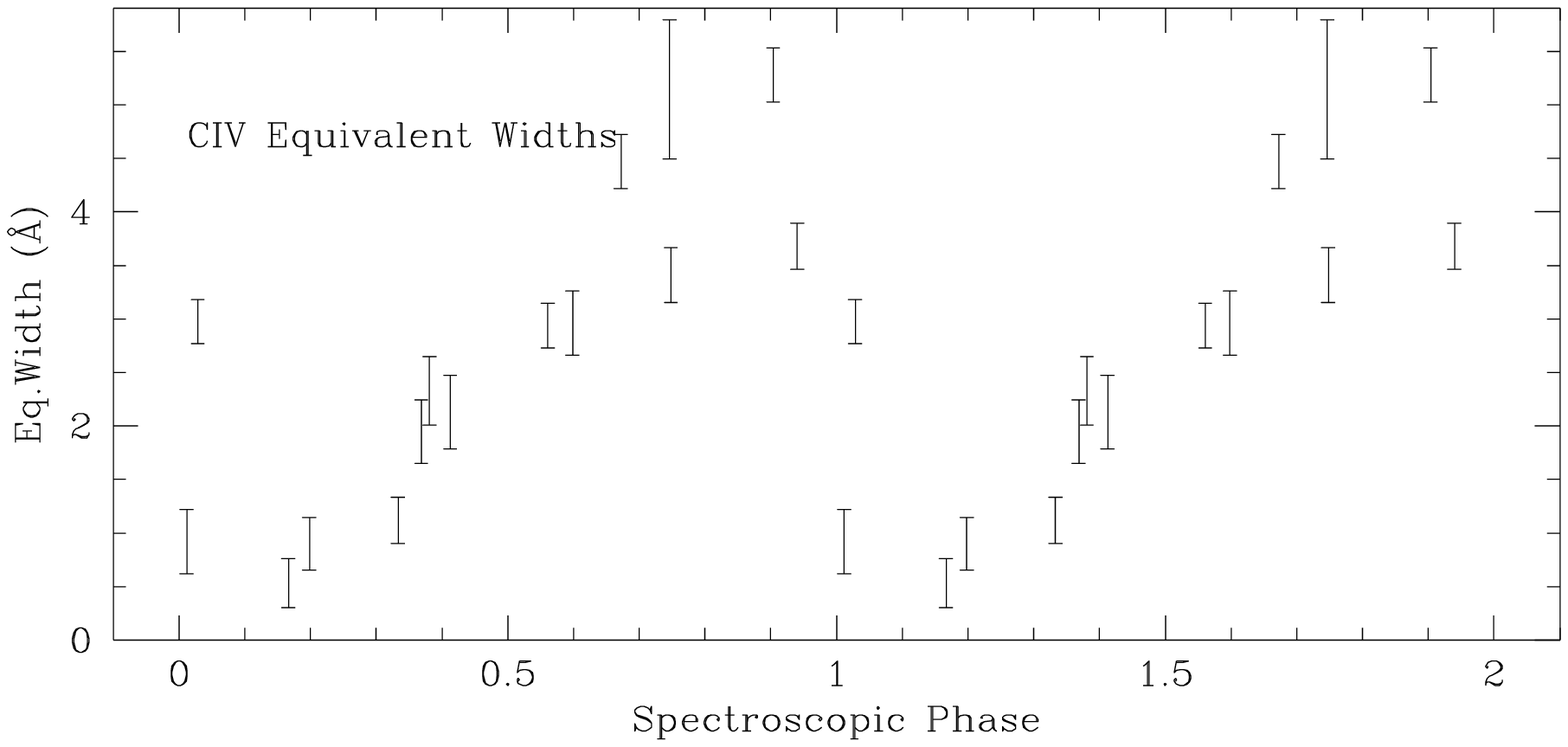}{4.0cm}{0}{65}{60}{-220}{-116}
\caption{TT Ari equivalent widths from CIV absorption.}
\label{fig-3}
\end{figure}

The resulting unabsorbed X-ray luminosity in the observed energy range
is $\rm{L_x}$(0.2-37.4 keV) = $1.4 \times 10^{32}$ erg $\rm{s^{-1}}$ assuming
an estimated lower limit distance of 200 pc (Shafter et al. 1985).
The observations of TT Ari suggest that neither ROSAT nor Ginga alone provided
enough information to adequately model its X-ray spectrum.  Fits to Ginga data
produce no evidence of the
softer component and over-estimate the absorption while ROSAT data alone places
only limited constraints on the bremsstrahlung continuum.
The energy range covered by ASCA, together with its spectral resolution
capabilities should greatly improve our understanding of
the multiple X-ray emission components in TT Ari and similar systems.

After years of observations,
the physical mechanisms producing variations in various spectral bands for TT
Ari are still not understood.  We began our analysis of the X-ray data by
comparing the observations
of TT Ari to similar systems.
Therefore, in Figure 2 we show the count spectra of TT Ari and three IPs
exhibiting some similar characteristics.
The general shapes of the count spectra are governed by the response of the
PSPC.  The systems
TV Col and TX Col have similar orbital and spin periods, but TV Col exhibits an
optical photometric
period slightly shorter than its orbital period and TX Col shows photometric
variations at the beat
period between orbital and spin periods.  TT Ari and V1223 Sgr have similar
orbital periods but V1223 Sgr
has a clearly defined spin period and photometric variations at the beat
frequency.  Initial spectral modeling
of the three IPs suggests a larger absorption column density exists for these
three
objects than for the harder spectral source TT Ari.  The IP spectra appear to
require multiple emission
components for fits to their spectra with fits in TV Col and TX Col suggesting
a combined soft optically thin plasma and harder bremsstrahlung continuum.

X-ray flares were detected in the simultaneous ROSAT and Ginga data sets but
were not found to be strictly periodic.  Simultaneous optical and X-ray flares
were also observed and analysis of the combined data sets continue.
Archival IUE data were extracted from observations obtained by C. Mansperger
during a portion of the simultaneous campaign
and observations by Verbundt 47 days later.  Variability
in the CIV ($\lambda1549$) line of TT Ari was first reported by Guinan \& Sion
(1980) who showed that
this line exhibits a P Cygni profile with great variability in both the
absorption and the
emission components.  A plot of equivalent width against spectroscopic phase
for the 1991 data
(using the Thorstensen et al. (1985) ephemeris) is shown in Figure 3.  The
absorption component
appears modulated on the spectroscopic period with maximum absorption occurring
near 0.8 phase.
Analysis of the variations on the reported photometric periods do not show such
a relationship.
The 47 day time span between the two IUE data sets significantly offsets the
photometric and spectroscopic phases to enable us to distinguish between
variations on these different periods.  The orbital
variability in the CIV absorption may be due to an asymmetric wind arising from
the disk or regions on the white dwarf.
However, the
variations in the emission component appear more complex
and may be independent of the source of absorption component variability.  We
are currently studying the relationship
between flaring events in X-rays and the variations observed with IUE.

\section{DP Leonis}
The eclipsing magnetic binary DP Leo belongs to the class of
CVs known as polars.
AM Her binaries, or polars, contain a synchronously rotating white dwarf with a
sufficiently
high magnetic field to disrupt the formation of a disk by material accreting
from its low mass companion. These systems,
reviewed recently by Cropper (1990) and Voikhanskaya (1990), and within the CV
review of C\'ordova (1993), exhibit cyclotron emission in the optical and IR,
soft
X-ray quasi-blackbody emission \mbox{(kT $\sim$ tens} of eV), and a hard X-ray
bremsstrahlung spectrum (kT $\sim$ tens of keV) with strong iron
emission lines superimposed. An accretion stream, accelerated to supersonic
speed, moves along the magnetic field lines of the white
dwarf toward one (or more) of the magnetic poles where a shock region, near to
the surface, is formed. Heated to tens
of keV, the post-shock matter cools through bremsstrahlung emission, while
electrons spiraling in the magnetic field (tens of
MG) produce the observed cyclotron emission. Reprocessing of the hard X-ray
photons in the stellar atmosphere
of the white dwarf
and/or large blob accretion in the white dwarf's photosphere produce the softer
X-ray component visible in all
 polars. The ratio of the
hard to soft X-ray fluxes in these systems provides important clues in
determining the sources of these X-rays.

\begin{figure}[hb]
\plottwo{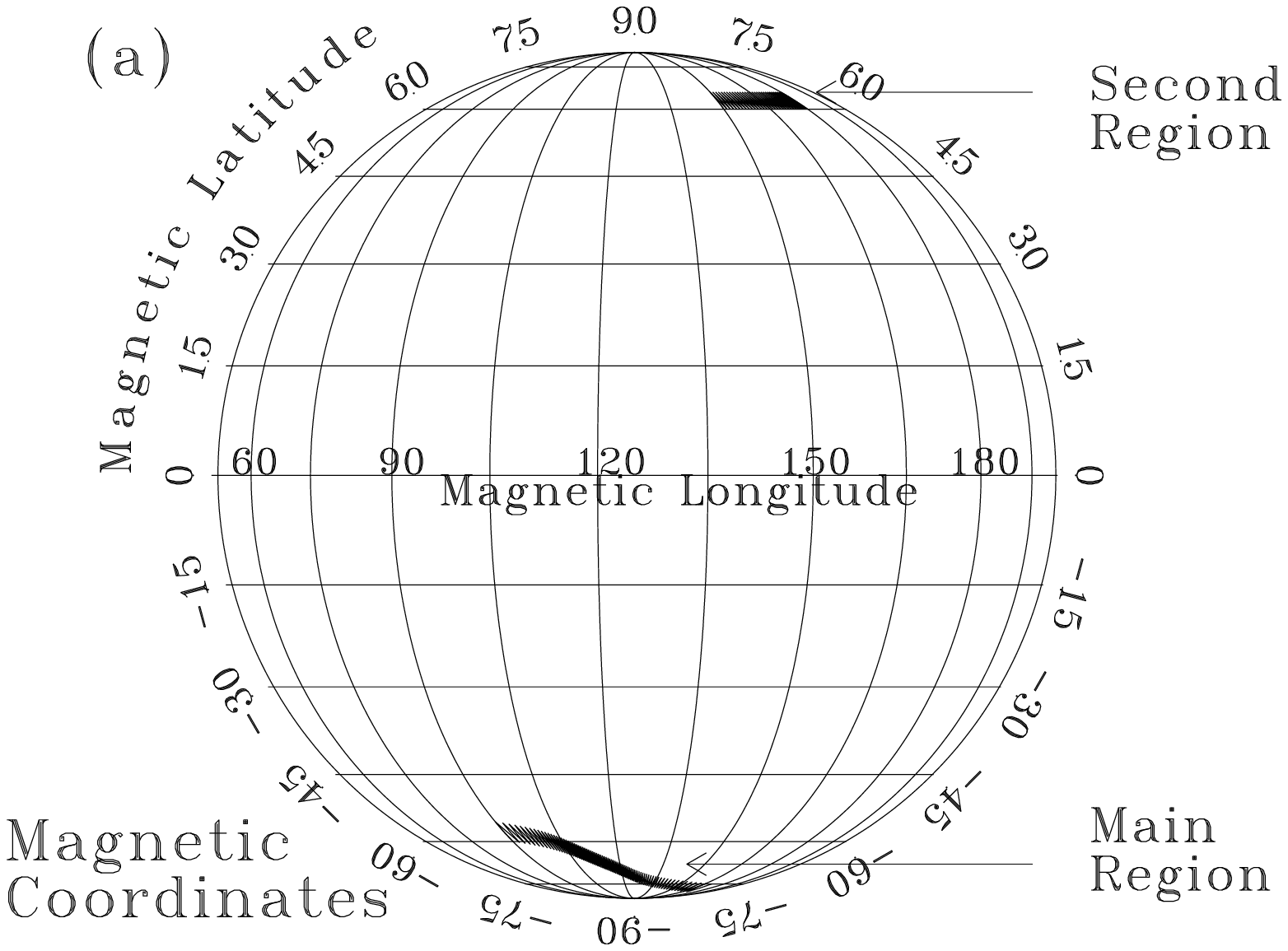}{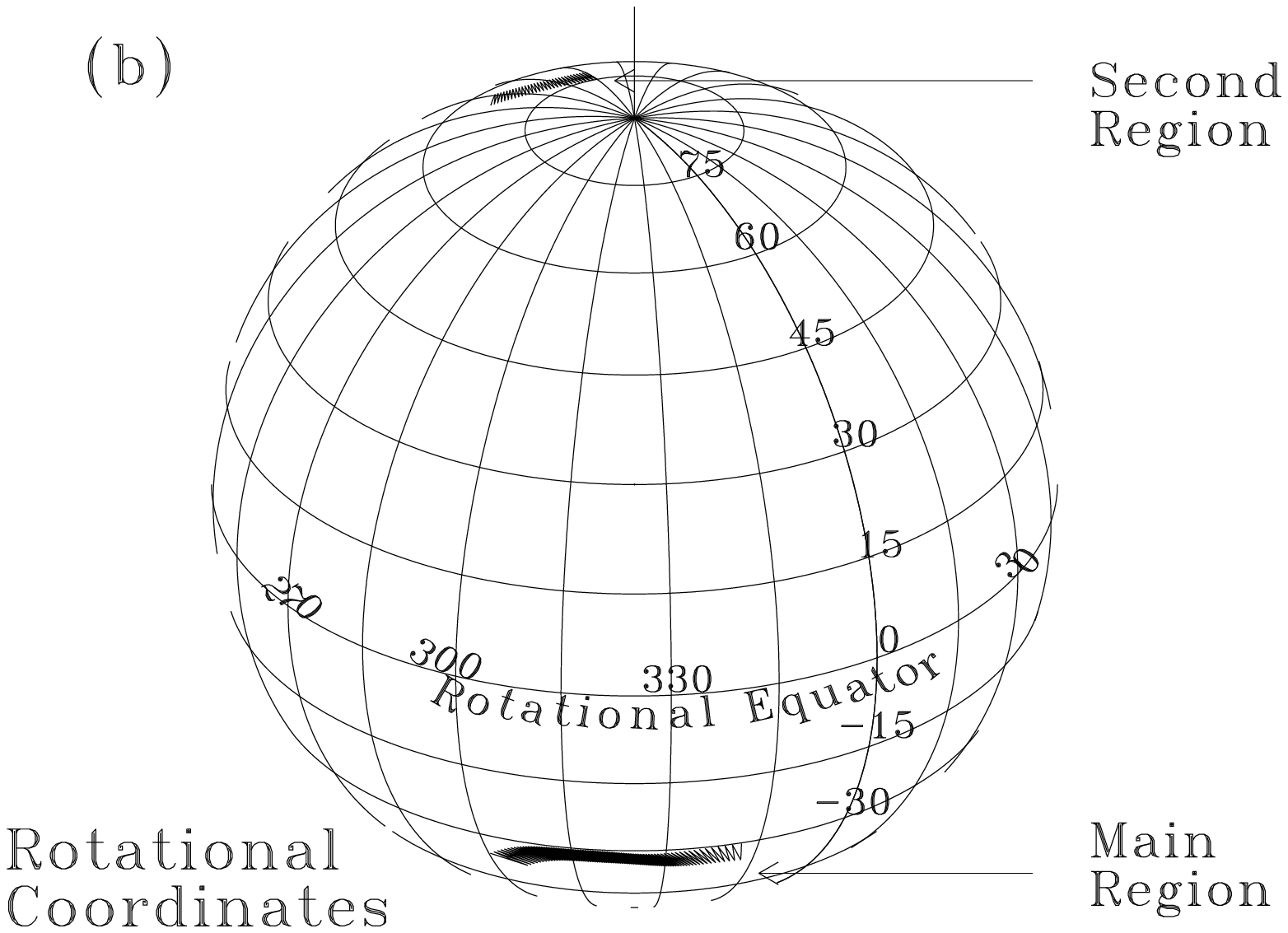}
\caption{Projections of the cyclotron emission regions for DP Leo as determined
by CW in (a) magnetic coordinates and (b) rotational coordinates.  Photometric
modeling suggests that these cyclotron regions and EINSTEIN observed X-ray
emitting regions may be coincident.  These diagrams correct the inconsistent
longitude scale within CW.}
\label{fig-4}
\end{figure}

Recent photometric observations by Bailey et al.\ (1993, hereafter BWF) suggest
that the longitude of the accretion stream impact spot
in DP Leo has changed from $-14^{\circ} \pm 4^{\circ}$ (1984) to $+4^{\circ}
\pm 3^{\circ}$
(1992), where a positive accretion spot longitude implies that the spot leads
the
secondary component.  A similar behavior of a variable accretion spot
position relative to the line of centers between the two components was
also shown to occur in another polar, WW Hor, but the variations occurred
in the opposite direction in longitude.  This opposite behavior together
with a clustering of accretion spot longitudes in polars near $+20^{\circ}$ led
BWF to conclude that
non-synchronous rotation of the white dwarf, which would not produce such a
skewed distribution, is an unlikely explanation for the observed accretion
spot longitudinal drifts.

The modeling, in DP Leo, of a strong cyclotron hump in phase resolved spectra
by Cropper \& Wickramasinghe (1993, hereafter CW) has
shown two cyclotron
emitting regions exist.  The main emitting pole is modeled by CW using a field
of 30 MG and is located between $30^{\circ}$
and $40^{\circ}$ below the rotational equator, trailing slightly the line of
centers of the two stars.  The
second emission region has a magnetic field strength of 59 MG, nearly twice
that of the main pole, and is positioned near the upper rotational pole but
nearly opposite to the secondary star (see Figure 4).  The differing magnetic
field
 strengths of the two poles and their angular separation of significantly less
than $180^{\circ}$ implies
a magnetic field configuration significantly more complex than a centered
dipole.  Wu \& Wickramasinghe (1993)
suggest a combination dipole-multipole model for polars where the white dwarf's
main pole oscillates
with respect to the secondary.

ROSAT PSPC observations were obtained of DP Leo during four intervals between
1992 May 30 and June 1 totaling 8910 seconds.
The observation intervals covered three eclipses of the X-ray emitting region
by the secondary.  The orbital period of DP Leo ($\sim$89 min.) is slightly
shorter than the orbital period of the ROSAT spacecraft which allowed for the
observation of the last two eclipses to be consecutive while the first eclipse
was observed one day earlier.
Photons detected within the acceptable time windows were extracted from a
circular
region centered on DP Leo with a radius of 3.5 arcminutes.  Five nearby
regions, each with the same radius as the source region and
apparently free of sources, were used to
determine the background level.  The background subtracted
source count rate was 0.27$\pm$0.01 cts $\rm{s^{-1}}$ over the entire
observation
and \mbox{0.34$\pm$0.01 cts $\rm{s^{-1}}$} within 0.3 phase of the eclipse (the
``bright phase'').

The spectral analysis was performed
using the ``DRM 36'' response matrix
which is the appropriate choice for AO2 observations (Turner \& George 1993).
The background subtracted source spectrum was binned
from 256 energy channels into the SASS standard 34 energy channels.
The  highest two energy channels were excluded
from further analysis due to uncertainties in the detector response.  The
lowest 4 out of the 34 energy
channels are known to be problematic due to the incorrect position
determination
of some low energy X-ray photons producing ``electronic ghost
images'' (Nousek \& Lesser 1993).
The effect of the soft energy halo created by these ghost images is greatest in
the first two energy channels and was
expected to be minimized in our analysis by extracting the source using a large
radius
(3.5 arcminutes).  However, a large drift in the energy to PI channel
transformation, calibrated at 1.49 keV, appears to have caused a deficit of
photons in the lowest two energy channels (Turner \& George 1993).  This effect
is seen in the ROSAT spectrum of DP Leo (Figure 5) where the steep initial
increase in the spectrum, visible in the first 3 channels, is larger than is
plausible for the energy resolution of the PSPC detector.  Therefore, while the
DP Leo spectrum is quite soft with no significant
detection of source photons in energy channels 12 or above (E $>$ 0.47 keV), we
determined it necessary to exclude the
first 2 energy channels from spectral fits.

The soft X-ray component in the spectra of polars is thought to arise from
either
large accretion column blobs which permeate the shock region above the
white dwarf and reach the photosphere where they cool through thermalized soft
X-rays, or through the reprocessing of hard X-ray bremsstrahlung emission,
produced in the post-shock region, and cyclotron radiation, generated by free
electrons flowing along the magnetic field lines (Frank et al. 1988).
The existence of a hard X-ray component in DP Leo cannot be confirmed in the
ROSAT data.

\begin{figure}[hb]
\plotfiddle{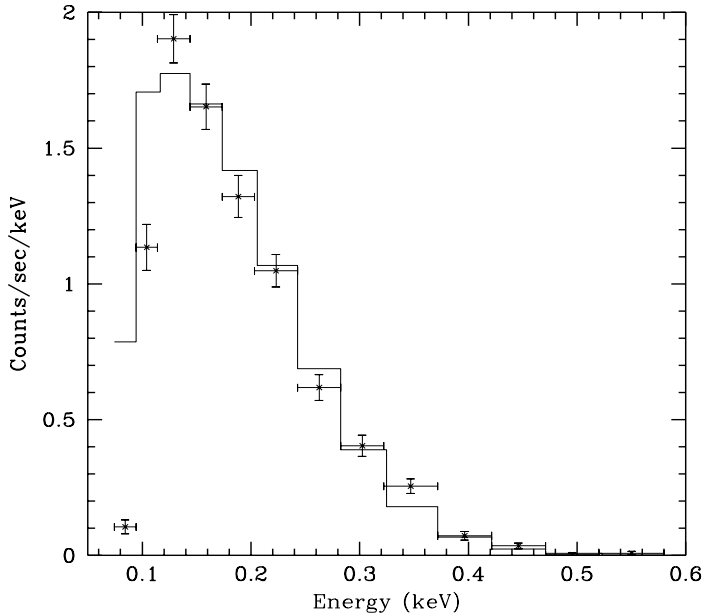}{7.2cm}{0}{55}{52}{-170}{-115}
\caption{ROSAT spectrum of DP Leo with blackbody fit. The first two channels
were not used in determining this fit (see text).}
\label{fig-5}
\end{figure}

During the 1992 ROSAT observations, 99 counts were detected in the source
region, above 0.47 keV, compared with a corresponding background of
\mbox{$74.8\pm10.1$} counts.  An upper limit to the
source intensity contributing the observed counts was calculated following the
Bayesian statistical
method and assuming a constant prior (Kraft, Burrows \& Nousek 1991).  The 90\%
confidence interval upper
bound is 41.3 counts resulting in a count rate of \mbox{4.64 x $10^{-3}$ cts
$\rm{s^{-1}}$} attributable to the source.

The observed soft spectrum was fitted using single component models with
absorption.
Acceptable fits were obtained both with a power law and with a blackbody
component.
However, physical models predict blackbody-like emission to occur.
For the
power law model, the best fit was found using an unabsorbed power law with
spectral index, $\alpha$, of 4.0.  The resulting $\chi^2$ statistic for
this model was 23.5 (28 d.o.f).  The blackbody model exhibited a best fit,
\mbox{$\chi^2$ = 25.3} (28 d.o.f.), where again the source was unabsorbed with
a
temperature of kT=$24.8^{+1.9}_{-5.4}$ eV (see Figure 5).  The calculation of
90$\%$ confidence intervals
limits the absorption column density to $\rm{N_H}$ $<$ 5 x $10^{19}$
$\rm{cm^{-2}}$ and allows
for blackbody temperatures between 19.4 and 26.7 eV (see Figure 6).

EINSTEIN observatory observations of DP Leo from 1979 (see Biermann et al.
1985) were extracted
and a spectral fit was performed to the data.  The spectral models were not
well constrained due to
the low sensitivity of EINSTEIN at soft energies.  However, the data were
consistent with the ROSAT
spectral fits and a harder component was not found.  EXOSAT observations of DP
Leo were obtained
in 1984 with a firm detection in the low energy (LE) detector using the
thin lexan filter but no source was detected using the aluminum/parylene filter
or with the
medium energy (ME) instrument (Schaaf et al. 1987).
\begin{figure}[hb]
\plotfiddle{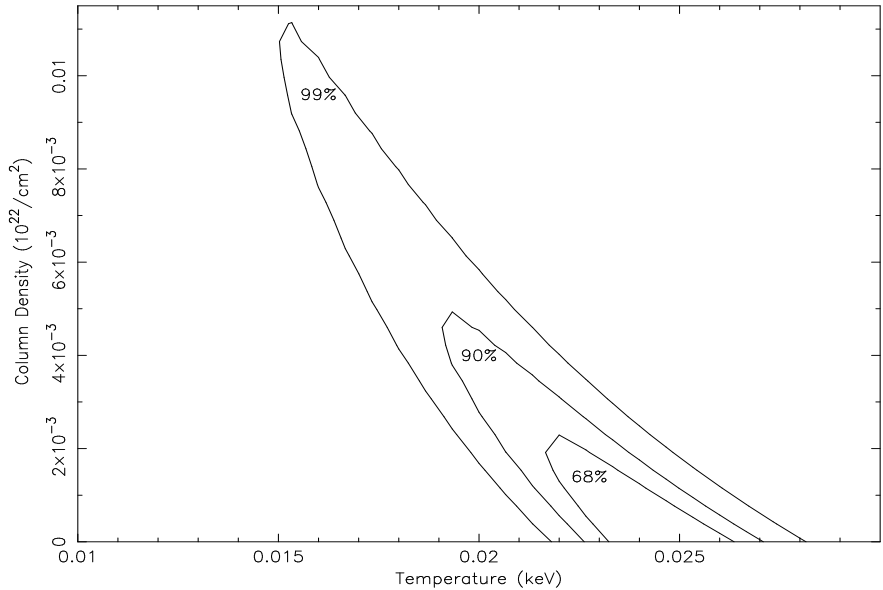}{6.7cm}{270}{50}{45}{-200}{248}
\caption{ROSAT blackbody fit confidence contours.}
\label{fig-6}
\end{figure}

A lower distance limit of 380 pc was previously determined
by Biermann et al. (1985) based upon their mass estimate for the secondary of
around $0.098 M_{\sun}$ and
their non-detection of the secondary to an R-magnitude of 22.  A measurement of
the R-magnitude
of the secondary and its mass was recently performed by BWF from a fit to their
photometric light curve.  They
obtained a solution with a mass of 0.106 $M_{\sun}$ for the secondary and an
R-magnitude of 21.8.  Based upon the work of
Young \& Schneider (1981), the absolute R-magnitude of such a star should be
around 13.54.  The
system has little absorption resulting in a distance estimate of 450 pc.

An independent upper limit to the
distance is obtained from the X-ray spectral fit.  The DP Leo system is located
in a region of
unusually low density. The distance could be as
far as approximately 500 pc, based upon the observations of this region by
Frisch \& York 1983, while still exhibiting a column density \mbox{$<$ 5 x
$10^{19}$ $\rm{cm^{-2}}$} (the upper limit
in the 90\% confidence contour).  The derived blackbody spectral fit together
with the calculated distances, when combined with the appropriate geometrical
corrections based upon viewing angle,
yield an upper limit to the source luminosity of $\rm{L_x}$(0.1 - 0.5 keV) =
6.0 x $10^{31}$ erg $\rm{s^{-1}}$ for a distance of 500 pc and an estimate of
$\rm{L_x}$(0.1 - 0.5 keV) = \mbox{4.8 x $10^{31}$ erg $\rm{s^{-1}}$} for a
distance of 450 pc.

The X-ray light curve from the PSPC, derived using all 34 energy channels, is
presented in the top plot of Figure 7 for
the entire data set and the lower plots for each of the individual eclipses.
The
phase was computed using the ephemeris of Schmidt (1988): HJD2444214.55283(17)
+ 0.062362849(6)E.  The data are
binned into 200 equally sized bins of length 26.94 seconds each.  Two dips
are visible in the light curves with the first dip centered near phase 0.943.
The second dip, centered slightly
before 1.0 phase,
is suspected of being the previously observed
eclipse of the X-ray emitting region on the white dwarf
by the secondary companion.  No evidence exists for accretion onto the
second pole, as was suggested by EXOSAT observations from 1984 (Schaaf et al.
1987) and
by cyclotron humps in phased resolved spectra from 1988-89 (CW).

\begin{figure}[hb]
\plotfiddle{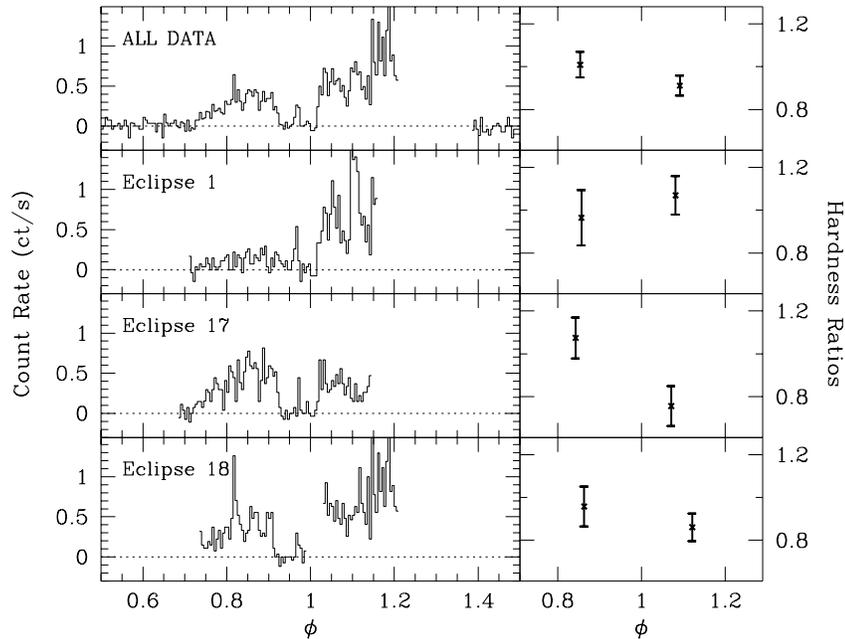}{7.3cm}{0}{55}{55}{-170}{-118}
\caption{ROSAT light curves and hardness ratios for DP Leo.}
\label{fig-7}
\end{figure}

Hardness ratios determined before and after eclipse are shown on the right side
of Figure 7 for each
of the ROSAT light curves.  The ratio was defined as PI channels 18 through 46
(0.18 - 0.47 keV) divided by PI channels
8 through 17 (0.08 - 0.18 keV) in the 256 channel system.  This definition
allowed for a hardness ratio near unity for the
entire observing set.  The ROSAT light curves are clearly variable within a
particular orbit and between orbits.  The
first observed eclipse showed little emission prior to the first intensity dip.
 The spike between the suspected position
of the intensity dip and the eclipse was evident and strong variable X-ray
emission was observed after eclipse.  The
next observed eclipse, Eclipse 17 (16 DP Leo orbits later), showed stronger
emission before the intensity dip than after the eclipse.
The final eclipse observation, Eclipse 18 (only one DP Leo orbit later), showed
emission before the intensity dip
similar to Eclipse 17, and flaring behavior after eclipse similar to Eclipse 1.

A comparison of the
dearth of emission prior to eclipse in Eclipse 1 to the much stronger emission
in Eclipse 17 shows that the hardness ratio
likely stayed around the same or increased slightly.  If the change
were caused by photoelectric absorption alone, a decrease in the hardness ratio
would have been expected.
Data after eclipse in Eclipse 17, which shows a dearth of flares, imply a
softer spectrum when compared to both data after eclipse in Eclipse 1 and data
before eclipse in Eclipse 17 where emission variations appear more active.
However, inferences based upon hardness ratio variations are limited due to the
paucity of observed photons.

\begin{figure}[hb]
\plotfiddle{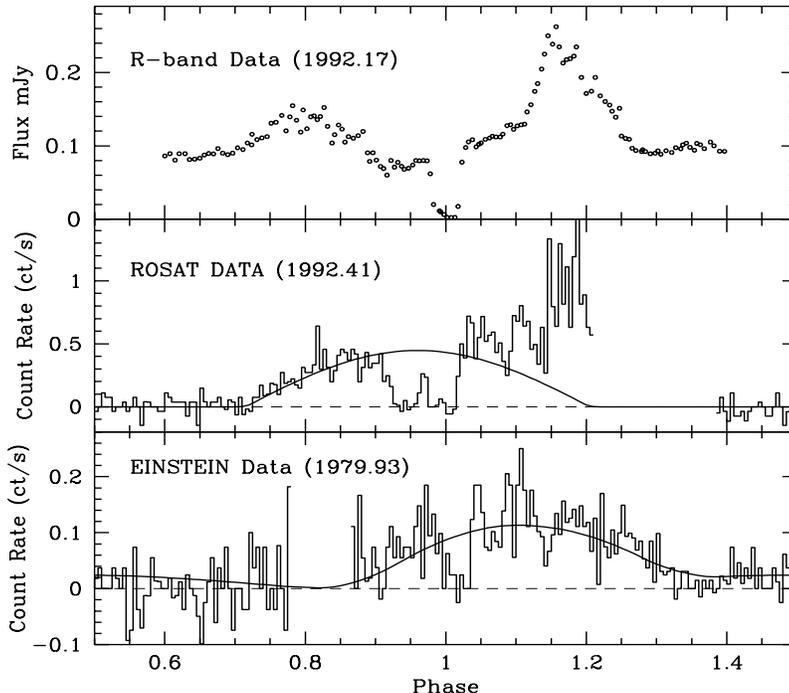}{8.05cm}{0}{55}{50}{-170}{-98}
\caption{DP Leo light curves in R-band (from a plot by BWF), ROSAT and
EINSTEIN.  The dashed lines correspond to count rates of zero.  The solid lines
are model fits to the data.}
\label{fig-8}
\end{figure}

High precision
optical photometry of DP Leo taken less than three months before the PSPC
observations show the center of eclipse to occur 24$\pm$4 seconds prior to the
prediction of the
Schmidt ephemeris (BWF; Bailey 1993).  The duration and timing
of the optical eclipse agree well with the interpretation of the second X-ray
dip as originating from the eclipse by the secondary (see Figure 8).

The duration of the X-ray eclipse in DP Leo is 3.6$\pm$0.3 minutes with the
center
occurring at phase 0.996$\pm$0.002 phase (all phases mentioned are according to
the Schmidt ephemeris) using the
combined data set.  Times of eclipse centers were calculated for the two
observed eclipses with complete coverage.
We obtained two minima timings at HJD2448773.21442$\pm$0.00012 and
HJD2448774.21235$\pm$0.00012.  A
revised ephemeris is then obtained
using the above timings together with those listed by Biermann et al. (1985),
Schaaf et al. (1987), Schmidt (1989),
and Bailey (1993) with each timing weighted according to its associated error:

Min. = HJD2444214.552934 + 0.0623628437E

    \hskip3.06cm $\pm$0.000043    \hskip0.1cm  $\pm$0.0000000007

The X-ray emission observed in DP Leo is assumed to be emitted as an optically
thick blackbody modulated by the
cosine of the viewing angle.  Photometric modeling was performed on the system
allowing for modulation
based upon viewing angle effects and occultation of the emission region as it
rotates out of view.  The geometry of the eclipse by the secondary was not
included and data within the eclipse were excluded when fitting models.  The
initial
sizes and locations of the X-ray emission regions were chosen to coincide with
the cyclotron emission regions found by CW.  The fluxes of these two
regions as well as the sizes and positions of the regions may be fixed,
constrained between limits or allowed to vary freely while the
Levenberg-Marquardt method of nonlinear least-squares was used to fit the model
to the observed X-ray light curves.

The EINSTEIN
light curve was well fit by constraining the X-ray emission regions to the
positions determined for the cyclotron emission
regions. The fluxes of the emission regions were the only free parameters used
to produce the model to the light curve shown in the EINSTEIN data set plot
of Figure 8.  The main region dominates the X-ray flux in the system from near
phase 0.85 out to near phase 1.35 where the secondary
emission begins to dominate.  Unlike the observations with ROSAT, the level of
X-ray emission is clearly above zero in EINSTEIN from
phase 0.4 to near 0.55 indicating the presence of either a greatly extended
main emission region or an additional second region.  Optical observations from
1982 to 1989 (Biermann et al. 1985, CW) show that accretion was occurring at a
second pole and supports the likelihood of two pole accretion also existing at
the time of the EINSTEIN observations.
Our photometric model suggests that the flux in EINSTEIN counts per unit area
of the secondary accretion region is 55\% of the main accretion region in
X-rays observed by EINSTEIN.

The X-ray emission in 1992 from ROSAT, however, shows no evidence of X-ray
emission from a second pole.  The R-band photometry
of BWF from 1992 (see Figure 8)  was fit by BWF with a single circular emission
region centered at or near the rotational meridian with
a radius of 0.15 times the white dwarf's radius.  The assumption of a
coincident X-ray emitting region will not reproduce
the observed ROSAT light curve.  The data from phase 0.4 to the intensity dip
were fit by a single circular emission region allowed to vary in
rotational longitude, flux and size.  The rotational colatitude of the spot was
fixed at the BWF determination of $100^\circ$.  The resulting model is shown in
the ROSAT data set plot of Figure 8.  The longitude of the spot is
$14^{\circ}_{\cdot}4$ and the spot radius is $3^{\circ}_{\cdot}2$.  The model
does not reproduce the ROSAT observations after eclipse which may be dominated
by a variable rate or size of
accretion blobs producing numerous flaring events.  The substantial amount of
emission still observed near phase 1.2 suggests
that the emission may be occurring over a large region, a region with
significant vertical extent or perhaps several distinct regions.

An estimate of the longitude of the X-ray emitting region may be obtained using
two
methods.  The first method uses the intensity dip prior to the eclipse.  The
intensity
dip lasts for a duration of 3.8$\pm$0.3 minutes with a center at
0.943$\pm$0.002 phase in the
Schmidt ephemeris.  Under the assumption that the intensity dip is an
occultation of
the X-ray emitting region by an accretion column aligned radially near the
white dwarf surface,
the longitude of the emitting region would be
$18^{\circ}_{\cdot}9\pm2^{\circ}_{\cdot}5$.

The second method is to measure the midpoint of the bright phase emission.  The
bright
phase is estimated to start near phase 0.72.  However, the end point was not
observed.
If the intensity dip center corresponds to the longitude of the emitting
region, then the bright phase would be expected to end near phase
1.16.  Data extend through this phase with no end to the bright phase observed
before phase 1.21 and a limit to the end of the bright phase exists at 1.39
when data again resume.  This
implies that the longitude of the accretion region center lies between
$-20^{\circ}$ and $+12^{\circ}$.  The allowed range is in conflict with the
first method.

This conflict may be resolved in several ways.  One resolution requires that
X-ray absorbing matter
lie along the line of sight to an X-ray emitting region positioned on the white
dwarf but in a position consistent with the
second method's allowed positions.  Such material would be required to be
stable in its relative
position, in both the orbit of DP Leo and in the size of the structure, over a
time period of at least one day.  In addition, it
must allow for less of an obscuration
of the emission region just prior to the eclipse by the secondary.  A possible
source for such emission could be a
solar-like prominence on the surface of the secondary star.  Assuming the
system parameters for DP Leo
fit in the paper by BWF together with an emission region located at $0^\circ$
longitude, such a region would
need to extend around 4 x $10^4$ km, or around 0.4 times the radius of the
secondary star, above the surface of
the secondary.

\begin{figure}[h]
\plotfiddle{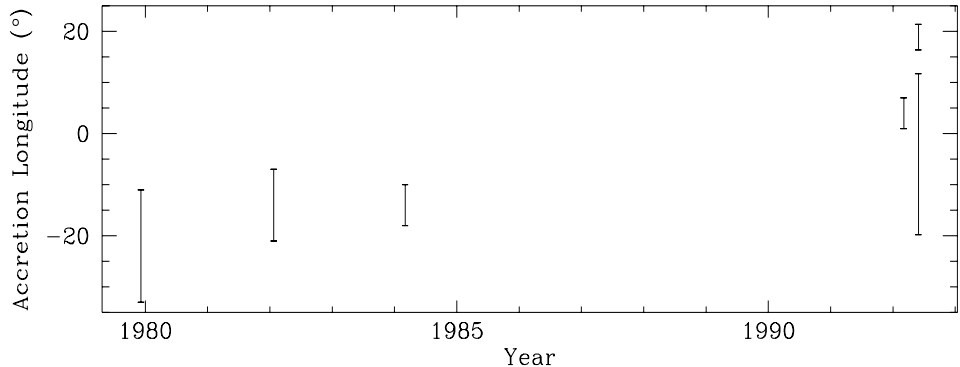}{3.9cm}{270}{50}{55}{-197}{305}
\caption{DP Leo accretion spot longitude positions v. time.}
\label{fig-9}
\end{figure}

Another solution is for the accretion stream to be shaped to completely block
the emission region at phase 0.943 and
partially just prior to eclipse, yet have the emission region still close to
the line of centers.  A combined centered dipole and inclined quadrupole field
for the white dwarf may allow for such a geometry and explain some of the
observed features in DP Leo.  The
multipole field model of polars by Wu \& Wickramasinghe (1993) provides for
three magnetic poles on the surface of the white
dwarf with two narrow poles lying close to the rotational axis and a band-like
pole, of field strength
half the axial poles, near the rotational equator.  The quadrupole field of the
white dwarf, which
falls off by $R^{-4}$ with distance from the white dwarf, channels the
accretion stream onto the
equatorial pole only near to the white dwarf (within several white dwarf
radii).  The
accretion stream, outside of the range dominated by the quadrupole field, has a
motion dictated by the momentum of the
Roche lobe overflowing material and the dipole field of the white dwarf.
The position of the secondary region suggests that a centered dipole,
positioned so that one pole lies near the secondary region and the other
$180^\circ$ away, will tend to channel the stream ahead
of the main region and out of the orbital plane.  Close to the white dwarf, the
quadrupolar field dominates and distorts the
stream to the main accretion region.  The stream, therefore, could be leading
the main accretion region, as would be necessary
to produce an intensity dip at the phase observed, even though the main
accretion region is near the line of centers.  The
distortion of the accretion stream much closer to the white dwarf, produced by
the quadrupole field, may then allow for a geometry where only partial
occultation of the
main emission region is observed just prior to eclipse.

The variations in accretion spot longitude as a function of time are presented
in Figure 9 from observations by
Biermann et al. (1985), BWF and the present work.  Both the ROSAT intensity dip
and allowed bright phase ranges are shown.  The determination of the center of
the bright spot region ($4^\circ \pm3^\circ$) by
BWF using their blue filter data would not appear to coincide with a similar
determination using their
red data (shown in Figure 8 of this paper) which would suggest a slightly
larger accretion spot longitude.  Evidence of an intensity dip around the phase
determined with ROSAT was presented by Schaaf et al. (1987) for one out of two
observations by EXOSAT from 1984.  This may indicate that the geometrical shape
of the absorbing material may be stable on time scales of hours to days (from
ROSAT observations), variable on time scales of months (from the EXOSAT
observations), yet recur after several years (between EXOSAT and ROSAT
observations).  However, the EXOSAT data are complicated by low counts, a high
background and the flaring of nearby X-ray sources.

\section{Conclusions}
We have determined X-ray spectral models and studied the photometric behavior
from observations of TT Ari and DP Leo.
TT Ari continues to show great variability in hard and soft X-rays as well as
in the optical and UV.  Flaring in the optical and X-ray data appear well
correlated.  This implies that efficient energy transfer, perhaps acoustic and
magnetohydrodynamic energy transport between a disk and its corona, exist
between the sources of these emissions.  The relationship of the UV emission to
the physical processes producing the variability at other wavelengths is
currently unclear.

X-ray variability in DP Leo
constrains the location of emission regions and absorbing material.
X-ray observations provide a useful mechanism for the determination of
variations in
accretion spot positions and can, therefore, serve to test models of
synchronism for the
rotation of the white dwarf or movement of the magnetic field on the white
dwarf.
Combined dipole/multipole models of the magnetic fields in polars may be
constrained in the future by determining the shape of
the accretion flow close to the white dwarf surface perhaps through combined
X-ray and cyclotron emission observations
of high inclination systems.

Finally, recent arguments by Shu (1993) suggest that classical T Tauri systems
may contain a disk similar in structure to current theories on
IPs where the inner disk is disrupted by the central star's magnetic field
The wind observed in these systems is then produced in the inner portion of the
remaining accretion disk. The application of some of the physical models
associated with CVs may find
more use in the study of the mass accreting T Tauri stars than previously
believed.  The reverse may also be
true.

We thank M. Cropper, J. Nousek, R. Thompson and R. Wade for helpful
discussions, J. Bailey for providing information on the
times of minima of the 1992 optical eclipses of DP Leo prior to
their publication and M. Ishida for his work and expertise on the analysis of
the Ginga data.

\end{document}